# ON MULTI-VIEW LEARNING WITH ADDITIVE MODELS

By Mark Culp, George Michailidis[1] and Kjell Johnson

*West Virginia University, University of Michigan and Pfizer Global Research & Development*

In many scientific settings data can be naturally partitioned into variable groupings called views. Common examples include environmental (1st view) and genetic information (2nd view) in ecological applications, chemical (1st view) and biological (2nd view) data in drug discovery. Multi-view data also occur in text analysis and proteomics applications where one view consists of a graph with observations as the vertices and a weighted measure of pairwise similarity between observations as the edges. Further, in several of these applications the observations can be partitioned into two sets, one where the response is observed (labeled) and the other where the response is not (unlabeled). The problem for simultaneously addressing viewed data and incorporating unlabeled observations in training is referred to as *multi-view transductive learning*. In this work we introduce and study a comprehensive generalized fixed point additive modeling framework for multi-view transductive learning, where any view is represented by a linear smoother. The problem of view selection is discussed using a generalized Akaike Information Criterion, which provides an approach for testing the contribution of each view. An efficient implementation is provided for fitting these models with both backfitting and local-scoring type algorithms adjusted to semi-supervised graph-based learning. The proposed technique is assessed on both synthetic and real data sets and is shown to be competitive to state-of-the-art co-training and graph-based techniques.

**1. Introduction.** In many scientific applications the available data come from diverse domains which are referred to as *views* henceforth. The views may consist of collections of numerical and categorical variables, but also may correspond to observed graphs. The objective of this study is to introduce a comprehensive modeling framework for a numerical or categorical response variable that is a function of data from distinct views. As

Received January 2008; revised July 2008.
[1]Supported in part by NIH Grant P4118627.
*Key words and phrases.* Multi-view learning, generalized additive model, semi-supervised learning, smoothing, model selection.

a motivating example, consider a collection of documents belonging to a particular scientific domain, for example, papers in statistics journals. The available information about the documents can be organized in the following three views: the corpus of the documents, that is, a collection of words in the documents [Blum and Mitchell (1998)]; information describing the documents (e.g., title, author, journal, etc.) [McCallum et al. (2000)]; and the co-citation network (graph) [McCallum et al. (2000), Neville and Jensen (2005)]. For the graph, nodes correspond to documents (observations) and edges count the number of citations to the same papers (pairwise similarity). The goal for this problem is to classify a document according to an attribute (e.g., whether a paper is applied or theoretical) where the attribute is known (labeled) for only a subset of the documents, with the remainder being unknown (unlabeled). In this context, the documents must be labeled by human action, whereas the view information can be obtained in an automated fashion (i.e., the set of labeled observations $L$ is significantly smaller than the unlabeled one; $|L| \ll |U|$). Further, it is worth noting that the first two views can be structurally represented by a data matrix with rows corresponding to observations (documents) and columns to variables, but the third view is given directly in the form of an observed graph.

Another example of multi-view data arises in drug discovery applications. Suppose that a very large number of characteristics (e.g., $> 1000$) has been collected for a library of chemical compounds. These characteristics range from high throughput screening measurements of compounds' effectiveness against numerous biological targets [Lundblad (2004), Hunter (1995)] to a compound's absorption, distribution, metabolism, excretion and toxicity (ADMET) properties [Fox et al. (1959), Kansy, Senner and Gubemator (2001)]. Further, given the chemical structure of a compound, it is nowadays fairly easy to computationally measure physical properties of each compound [Leach and Gillet (2003)]. Given data on the response of a subset of compounds in a library for a particular target (e.g., whether or not a side-effect is associated with the compound), the goal is to use the data available in these diverse views (biological, chemical, ADMET) to predict the response for the remaining members in the library. Notice also that the target status of a potential drug can be both time consuming and hard to determine (e.g., side effects in humans may take many years to appear), whereas the biological and chemical compound characteristics can be obtained in a shorter time period (usually days to weeks) and with less effort (hence, $|L| \ll |U|$). Other examples of multi-view problems are present in applications involving genomic [Nabieva et al. (2005)] and proteomic data [Yamanishi, Vert and Kanehisa (2004)].

As illustrated with these examples, the available data can be naturally partitioned into disjoint data sets, referred to as views, that in some cases can be represented as data matrices, while in other cases come in the form of graphs. Views comprised of variables can differ in the number of variables, variable type (numerical, ordinal, nominal), noise level and scale. Graph views may differ in the node degree distribution, type and distribution of edge weights. Traditionally, models for the prediction problem at hand have been built that include all the variables available, without taking into consideration the presence of distinct views. Further, data in the form of an observed graph were ignored completely. Popular techniques for building flexible prediction models include recursive partitioning [Breiman et al. (1984)], multivariate adaptive regression splines [Friedman (1991)], random forests [Breiman (2001)], support vector machines [Vapnik (1998)], partial least squares [Mevik and Wehrens (2007)], etc. Nevertheless, there are several situations where incorporating distinction of views offer a number of advantages from a data analysis point of view, including:

- *View level analysis*: In many applications it is of great interest to develop a model that provides insight into the underlying relationship among views, potentially identifying interactions between them, and also to assess their predictive capabilities. The latter can prove particularly useful in problems where collecting the necessary data for a view may be resource demanding and thus expensive.
- *Incorporation of graph information*: As already discussed, in many applications some of the available data come in the form of a graph that available statistical models can not handle in a straightforward manner.
- *Improving predictive performance*: Allowing the available data to be partitioned into different views and incorporating interactions among them offers the advantage of building more flexible and potentially more powerful models, exhibiting better performance in terms of prediction accuracy.

In this work we introduce an additive modeling framework that takes into consideration the presence of distinct data views. Further, it incorporates in a seamless fashion observed graphs, allows for view level analysis and on many occasions leads to significant gains in performance.

The main idea of this framework is to represent each view by a linear smoother. The difficultly is in providing representative multi-dimensional view smoothers on any data type (graphs or numerical/categorical), while accounting for the sparse labeling of the response which occurs in several of the applications under consideration ($|L| \ll |U|$). To define a smoother for an observed graph, we build on recent advances in graph-based transductive learning [Blum and Mitchell (1998), Zhu (2007)]. Specifically, graph-based transductive learning addresses the problem of learning in a setting where the available data come in the form of a graph (labeled and unlabeled observations correspond to vertices/nodes and pairwise associations to edges), where a numerical or categorical variable can also be associated with each node on the graph. In this context, Culp and Michailidis (2008a) note that the adjacency matrix of the appropriately normalized graph leads to a stochastic matrix that resembles a kernel smoother with a transductive form defined on both labeled and unlabeled nodes. In this work we define a *transductive smoother* in general as a linear smoother defined for a response that has a missing unlabeled component. In the case of numerical, categorical or ordinal data views, it is fairly straightforward to extend a classical linear smoother [see, e.g., Hastie and Tibshirani (1990)] into a transductive one [Culp and Michailidis (2008a)]. Upon obtaining the transductive smoother for each view, the next challenge is in fitting a model to a smoother of this form, since the smoother is linear in the response partitioned with a labeled (observed) and unlabeled (missing or unobserved) component. To address this, we propose a novel generalized fixed point self-training framework (Section 2) that essentially extends the classical generalized additive model into the multi-view transductive setting. Under reasonable conditions on the transductive smoother, the solution is guaranteed to uniquely exist. In addition, the computational issues are addressed using established iterative self-training procedures for both the regression and classification settings [Culp and Michailidis (2008a), Zhu (2007)].

The proposed modeling framework treats both the variable and graph views represented by the transductive smoothers as the equivalent of "variables" in a generalized additive model which can subsequently be fitted by an extension of the common backfitting (local scoring) algorithm to self-training. Due to the linearity of the solution in the response variable, existing model selection techniques can be readily applied to select important views. Also, the smoothers require estimation of underlying parameters, as in the classical case, and we investigate a criterion more appropriate for transductive smoothers defined on views. The results indicate that the multi-view model using this estimation approach is quite competitive with the state-of-the-art multi-view techniques discussed next.

1.1. *Relevant existing multi-view learning approaches.* We provide next a brief exposition of existing approaches geared toward improving accuracy in multi-view learning problems.

It is natural to consider the general semi-supervised classification problem as a precursor to the multi-view setting. In semi-supervised learning a relatively small percentage of the observations (cases) contain labels. The objective is to use the labeled cases and their relation to the unlabeled cases to complete the labeling of the data. Upon label completion, the classifier can be used to predict new cases (inductive) or must be retrained/updated (transductive) to incorporate this information into the classifier. Various algorithmic solutions available for this problem include self-training [Abney (2004)], graph regularization [Wang and Zhang (2006)], semi-supervised SVM [Chapelle, Sindhwani and Keerthi (2008)], and parametric models

MULTI-VIEW LEARNING WITH ADDITIVE MODELS 5

[Krishnapuram et al. (2005)]. For example, in Zhu, Ghahramani and Lafferty (2003) the authors propose a quadratic energy optimization problem leading to a harmonic estimate for the unlabeled data with the constraint that their labeled estimate retains the original labels. This approach has several connections with electrical circuits [Zhu, Ghahramani and Lafferty (2003)], ST-minicut clustering techniques [Blum and Chawla (2001)], spectral kernel techniques [Joachims (2003), Johnson and Zhang (2007)], and kernel smoothing approaches [Lafferty and Wasserman (2007), Culp and Michailidis (2008)]. The survey by Zhu (2007) and the book by Chapelle, Schölkopf and Zien (2006) highlight several of these semi-supervised approaches and address both theoretical and practical issues.

In multi-view learning Blum and Mitchell (1998) developed a co-training procedure for classification problems that is based on the idea that better predictive models can be found at the individual view level, rather than fitting a model directly on all the available views. The co-training procedure trains a separate classifier for each view and then proceeds in a self-training fashion by iteratively treating the most confident unlabeled observations as true labeled ones using the fitted class estimates as the true values. After a prespecified number of iterations, co-training produces a classification of every observation in the data. A final classifier can be formed through a combination of the individual view classifiers, in order to predict new observations unavailable during the training phase. The intuition behind this approach is that if the results of the individual classifiers arrive at the same classification for either a labeled observation (known response) or, more importantly, an unlabeled observation (unknown response), and the views are conditionally independent, then it is highly likely that the derived classification is correct [Blum and Mitchell (1998), Abney (2002)].

Another set of procedures are based on a transductive graph-based learning setting [Joachims (2003), Zhu (2007), Culp and Michailidis (2008)], where the underlying graph is either observed directly or constructed from the data matrix. In this setting, each view can be captured by a graph with its corresponding adjacency matrix, and the views are integrated by adding their adjacency matrices. For example, the Spectral Graph Transducer (SGT) treats the resulting graph as an energy network, where the labeled observations are positive or negative sources and the objective is to determine an optimal energy estimate for the unlabeled responses [Joachims (2003)]. An approach that shares some of the SGT's characteristics is the Sequential Predictions Algorithm (SPA), which forms the final graph by using graph theoretic operations such as unions and intersections [Culp and Michailidis (2008)]. This procedure employs a local kernel smoothing algorithm with a regularized extrapolation penalty that shrinks the estimates of unlabeled nodes farther away from labeled ones toward the class prior distribution. It can be seen that such graph-based procedures can naturally incorporate multiple views. However, views comprised of numerical variables must first be converted into graphs, which may not be the most effective way of representing the data. Further, high performance classifiers such as support vector machines or random forests cannot be used in this setting.

The proposed modeling framework shares some features with existing co-training and graph based approaches. However, by building a smoother for each view and then combining them through a generalized additive model, the proposed approach offers useful tools such as view level analysis and incorporation of graph terms, together with performance improvements. The remainder of the paper is organized as follows: In Section 2 we introduce the modeling framework and address estimation and model selection issues. Section 3 illustrates the model on a number of real and simulated data sets. Some concluding remarks are drawn in Section 4.

**2. Modeling framework for multi-view data.** For the problem at hand, let $Y$ denote the response variable of length $n = |L \cup U|$, partitioned into the set $L$ of labeled observations and $U$ of unlabeled ones (specifically $Y_U$ is missing with $|U| \geq 0$), that is, $Y = [Y_L' Y_U']'$. The available predictors can be partitioned into $q$ *distinct* views, where views may consist of variables, observed graphs or both. Views comprised of numerical, nominal or ordinal variables are represented by data matrices $X_\ell$ of size $n \times p_\ell$. It is assumed that a particular variable can only be present in one view. Each individual data matrix can be partitioned row-wise into two disjoint labeled and unlabeled sets: $X_\ell = [X_{L_\ell}' X_{U_\ell}']'$. Views can also correspond to observed graphs, $G_\ell = (N_\ell, E_\ell)$, with $N_\ell = L \cup U$ denoting the node set and $E_\ell$ the edge set. The similarity weighted $n \times n$ adjacency matrix $A_\ell$ for observed graph $G_\ell$ can also be partitioned in the following way:

$$A_\ell = \begin{pmatrix} A_{LL}^\ell & A_{LU}^\ell \\ A_{UL}^\ell & A_{UU}^\ell \end{pmatrix},$$

with $A_{LL}^\ell$, $A_{UL}^\ell$, $A_{UU}^\ell$ representing edges between labeled nodes, between labeled and unlabeled nodes and between unlabeled nodes, respectively.

As noted above, the response variable is partitioned into a labeled and unlabeled component, which induces the corresponding partitions to $X$ and $G$, respectively. The proposed modeling framework accommodates multiple views, as well as their interactions, as follows:

$$\eta = \alpha + \sum_i f_{j_i}(X_{j_i}) + \sum_i f_{\ell_i}(G_{\ell_i})$$
(2.1)
$$+ \sum_{i,i'} f_{j_i,j_{i'}}(X_{j_i}, X_{j_{i'}}) + \sum_{i,i'} f_{\ell_i,\ell_{i'}}(G_{\ell_i}, G_{\ell_{i'}}),$$

where $\eta = g(\mu)$ denotes the link function of the response $Y = [Y_L', Y_U']'$ for which $Y_U$ is missing, with $\mathbb{E}(Y) = \mu$, $\alpha$ is an intercept term, $\{f_{j_i}(\cdot)\}$ are smooth functions defined on the feature space $X$, and $\{f_{\ell_i}(\cdot)\}$ are smooth functions defined on the nodes of $G$ [Culp and Michailidis (2008a), Zhu (2005)].

The main difficulty with (2.1) stems from the transductive nature of the model, due to the presence of the missing response vector $Y_U$. To fit (2.1), we propose next a two stage optimization framework referred to as *generalized fixed point self-training*. For this approach, we must first define the *training response* as $Y_{Y_U} = [Y_L', Y_U']'$ with $g(Y_U) \in \mathbb{R}^{|U|}$ an arbitrary initialization. The training response is then employed in two stages, each discussed in detail next to obtain an estimate $\hat{Y} = [\hat{Y}_L', \hat{Y}_U']' = g^{-1}(\hat{\eta})$.

In the first stage, the training response $Y_{Y_U}$ is employed to determine an estimate for $\eta = \alpha + \sum f_i$ by solving

$$(2.2) \qquad \min_{f=[f_1',\ldots,f_p']'} L(Y_{Y_U}, g^{-1}(\eta)) + J(f),$$

where $L(y, f)$ is a loss function that increases as the deviance from $y$ and $f$ increases, and $J(f)$ is an appropriate penalty term on $f$. The key issue is that of existence and uniqueness of the resulting estimate $\hat{\eta}(Y_U)$ as a function of $Y_U$. From this perspective, it can be seen that the posited problem is a "supervised" one with respect to response $Y_{Y_U}$ and data $X, G$. As a result, there are a number of well-known approaches that lead to a unique solution including SVMs, logistic regression, additive models, neural nets, etc. [Hastie, Tibshirani and Friedman (2001)]. Upon completing the first stage an estimate $\hat{Y}_{Y_U} = g^{-1}(\hat{\eta}(Y_U))$ is obtained for the entire response vector $Y$ as a function of $Y_U$.

The second stage deals with the problem of optimally determining an appropriate value of $Y_U$ necessary for training purposes. It can be chosen as the solution to the following optimization problem:

$$(2.3) \qquad \min_{Y_U} (g(Y_U) - \hat{\eta}_U(Y_U))'(g(Y_U) - \hat{\eta}_U(Y_U)),$$

that is, the deviance between $g(Y_U)$ and $\hat{\eta}_U(Y_U)$ is minimized. A moment of reflection shows that the optimal $Y_U$ corresponds to a fixed point. Existence and uniqueness of the fixed point $\hat{Y}_U = g^{-1}(\hat{\eta}_U(\hat{Y}_U))$ are a key issue [Kakutani (1941)]. In several cases the solution can be obtained in a direct manner, given the form of $\hat{\eta}(\cdot)$. However, in other circumstances the fixed point solution must be approximated. One way to approximate is using Newton's method whose $k$th update step is given by

$$\hat{Y}_U^{(k+1)} = \hat{Y}_U^{(k)} - (I - \nabla g^{-1}(\hat{\eta}_U(Y_U))|_{Y_U=\hat{Y}_U^{(k)}})^{-1}(\hat{Y}_U^{(k)} - g^{-1}(\hat{\eta}_U(\hat{Y}_U^{(k)}))).$$

A key assumption is that the maximum eigenvalue of the gradient $\nabla g^{-1}(\hat{\eta}_U(\cdot))$ is less than one, which renders the corresponding map a contraction, thus guaranteeing the existence of a fixed point. By the derivative chain rule, this approach requires the gradient of $\hat{\eta}_U(\hat{Y}_U^{(k)})$ for each $k$, which can be computationally demanding to obtain. This motivates the following slower iterative self-training algorithm [Culp and Michailidis (2008a)]:

1. Initialize the unlabeled response vector, $\hat{Y}_U^{(0)}$ and get tolerance $\delta$.
2. Iterate until $\|\hat{\eta}_U^{(k)} - \hat{\eta}_U^{(k-1)}\| < \delta$
   (a) Solve (2.2) with response $Y_{\hat{Y}_U^{(k-1)}} = [Y_L', \hat{Y}_U^{(k-1)'}]'$ and data $\{X_\ell, G_\ell\}$ to get $\hat{\eta}^{(k)}$.
   (b) Set $\hat{Y}_U^{(k)} = g^{-1}(\hat{\eta}_U^{(k)})$.

Convergence of this algorithm provides an approximation to the fixed point defined in (2.3) by construction. Whenever there exists an initialization that results in local convergence of the above procedure, then the fixed point exists and is approximated by the algorithm. Moreover, if the algorithm converges globally independent of the initialization, then the fixed point is uniquely approximated by the procedure. The global convergence depends on the specific choices for $\{f_j(\cdot)\}$ [Culp and Michailidis (2008a)]. We provide below the details for fitting this procedure first for squared error loss and then for logistic loss when $\{f_j(\cdot)\}$ are estimated using transductive smoothers.

Before we discuss transductive smoothers, we briefly address the bias and variance of an estimate $\hat{f}$ resulting from the proposed fixed point self-training approach in the regression context. To begin, we consider first the supervised case with the goal of estimating a function $\widetilde{f}_L$ from data $(X_L, Y_L)$, where the response $Y_L$ is continuous. It is well known that the supervised error of $\widetilde{f}_L$ with respect to response $Y_L$ can be decomposed into bias, variance and irreducible terms [Hastie, Tibshirani and Friedman (2001)]. Recall that the semi-supervised problem using the fixed point self-training method arrives at an estimate $\hat{f} = [\hat{f}_L', \hat{f}_U']'$ with data $(Y_{\hat{f}_U}, X)$. From this, the error of estimate $\hat{f}$ with respect to response $Y_{\hat{f}_U}$ can be decomposed as

$$\mathrm{Error}(\hat{f}) = \sigma^2 + \mathrm{Bias}(\hat{f}_L \mid X) + \mathrm{Var}(\hat{f}_L \mid X) + \mathrm{Error}(\hat{f}_U),$$

where $\sigma^2$ is the irreducible error term, $\mathrm{Bias}(\hat{f}_L \mid X)$ and $\mathrm{Var}(\hat{f}_L \mid X)$ are the respective bias and variance of the labeled estimate relative to the true function conditioned on the full data $X$, and $\mathrm{Error}(\hat{f}_U) = 0$ by construction. The resulting supervised and semi-supervised bias and variance terms are given by

| *Training* | *Bias/Variance* |
|---|---|
| Supervised | Error $= \sigma^2 + \mathrm{Bias}(\widetilde{f}_L \mid X_L) + \mathrm{Var}(\widetilde{f}_L \mid X_L)$ |
| Self-training | Error $= \sigma^2 + \mathrm{Bias}(\hat{f}_L \mid X) + \mathrm{Var}(\hat{f}_L \mid X)$. |

Therefore, the self-training approach allows one to determine a labeled estimate, $\hat{f}_L$, that balances the bias/variance tradeoff by employing the entire $X$ information, whereas the corresponding supervised problem achieves a similar goal by only using the $X_L$ information. Naturally, this decomposition extends in the presence of graph views.

REMARK 1. The connection between the two stage optimization approach and the self-training algorithm reveals that this framework is a semi-supervised example of a block relaxation algorithm [see the discussion in Leeuw (1994)].

2.1. *Fitting the additive model in the regression context.* For ease of presentation, we study the simplest form of (2.1) using the fixed point self-training approach that combines a numerical/categorical feature set $X$ and a graph view $G$ for a continuous response $Y_L$. The resulting model is given by

$$(2.4) \qquad Y = \alpha + f_1(X) + f_2(G) + \varepsilon.$$

Our strategy in fitting this model is based on constructing transductive smoothers for both the $X$ and $G$ views. We provide next the details of such a construction and extend it to incorporate interactions among these two views. The implementation details are presented in Section 2.4.

To fit the function $\eta = \alpha + f_1(X) + f_2(G)$ under a squared-error loss criterion, we must minimize with respect to $f = [f_1' f_2']'$ the following:

$$(2.5) \qquad \min_f (Y_{Y_U} - \eta)'(Y_{Y_U} - \eta) + \sum_{j=1}^{2} \lambda_j f_j' P_j f_j,$$

where $P_j$'s are penalty matrices for each view, $\alpha = \bar{Y}_{Y_U}$, and $\lambda_j > 0$ the associated tuning parameters. For a graph view we choose $P$ as the combinatorial Laplacian operator, that is, $P = D - A$, where $A$ is the adjacency matrix and $D$ is its row-sum diagonal matrix. For $X$ data views, there are many choices, including generalized additive models, spline-based models, nonparametric models, etc. [Hastie and Tibshirani (1990)]. Below we provide more details on the penalty matrix for the graph case (Section 2.1.1) and for the feature $X$ data case (Section 2.1.2). No matter how it is obtained, each penalty matrix emits the following partition:

$$(2.6) \qquad P_j = \begin{pmatrix} P_{LL_j} & P_{LU_j} \\ P_{UL_j} & P_{UU_j} \end{pmatrix}.$$

In the above expression, the submatrix $P_{LL}$ captures associations between labeled portions of the data, while $P_{UL}$ and $P_{LU}$ labeled to unlabeled data associations, and finally $P_{UU}$ unlabeled to unlabeled ones. Each penalty matrix, $P_j$, is assumed to be positive semidefinite, which in turn implies that the problem in (2.5) is jointly convex. Therefore, one can solve it to obtain the following equations:

$$\text{(2.7)} \quad \hat{f}_\ell - S_\ell\left(Y_{Y_U} - \sum_{j \neq \ell} \hat{f}_j\right) = \vec{0} \quad \text{for } \ell = 1, 2 \text{ with } S_\ell = C(I + \lambda_\ell P_\ell)^{-1},$$

where $C = (I - \mathbf{1}\mathbf{1}'/n)$ is a centering matrix. This is clearly an extension of the Gauss–Seidel algorithm with response $Y_{Y_U}$ and smoothers $\{S_\ell\}_{\ell=1}^2$:

$$\text{(2.8)} \quad \begin{pmatrix} I & S_1 \\ S_2 & I \end{pmatrix} \begin{pmatrix} \hat{f}_1(X) \\ \hat{f}_2(G) \end{pmatrix} = \begin{pmatrix} S_1 Y_{Y_U} \\ S_2 Y_{Y_U} \end{pmatrix}.$$

The solution of the Gauss–Seidel algorithm is well known to take the form $\hat{\eta}(Y_U) = \alpha + \hat{f}_1(X) + \hat{f}_2(G) = R Y_{Y_U}$ with smoother $R$ independent of $Y_{Y_U}$ [Hastie and Tibshirani (1990)]. Therefore, the first stage in our model fitting strategy results in a linear fitting technique, $\hat{\eta}(Y_U) = R Y_{Y_U}$.

For the fixed point step (2.3), we need to define the class of *transductive smoothers* generated from $X$, $G$ or both as follows:

$$\mathcal{A}[\cdot] = \{S : S \text{ is an } n \times n \text{ linear smoother matrix}$$
$$\text{constructed from source such that } \rho(S_{UU}) < 1\},$$

where $S \in \mathcal{A}[\cdot]$ emits the partition

$$\text{(2.9)} \quad S = \begin{pmatrix} S_{LL} & S_{LU} \\ S_{UL} & S_{UU} \end{pmatrix},$$

and $\rho(\cdot)$ denotes the spectral radius of the matrix under consideration. The partitions in the transductive smoother correspond to the partitions in the penalty matrix given above, that is, (2.6). From the first optimization problem discussed above, we get that for any smoother $S$ the unlabeled estimate is given by $\hat{\eta}_U(Y_U) = S_{UL} Y_L + S_{UU} Y_U$. The optimization problem in (2.3) subsequently reduces to

$$\min_{Y_U} ((I - S_{UU})Y_U - S_{UL}Y_L)'((I - S_{UU})Y_U - S_{UL}Y_L),$$

and therefore, the condition that $\rho(S_{UU}) \equiv \rho(\nabla \hat{\eta}_U(\cdot)) < 1$ results in $\hat{Y}_U = (I - S_{UU})^{-1} S_{UL} Y_L$ as the unique fixed point. From this, in the case of a regression model a closed form solution can be obtained as

$$\text{(2.10)} \quad \begin{pmatrix} \hat{Y}_L \\ \hat{Y}_U \end{pmatrix} = \begin{pmatrix} S_{LL} + S_{LU}(I - S_{UU})^{-1} S_{UL} \\ (I - S_{UU})^{-1} S_{UL} \end{pmatrix} Y_L.$$

As expected, the resulting predicted responses are linear in the labeled data $Y_L$, that is, $\hat{Y} = M_L . Y_L$, with $M_{LL}$ and $M_{UL}$ the respective $|L| \times |L|$, $|U| \times |L|$ matrices identified in parenthesis in the above expression.

A special case arises when $f_1(X)$ is modeled by a linear model, that is, $f_1(X) = X\beta$. The additive model in (2.4) reduces to the following semi-parametric model:

$$Y = X\beta + f_2(G) + \varepsilon,$$

which is fit using transductive smoothers, $S_1 = H = X(X'X)^{-1}X'$ and centered symmetric smoother $S_2 = S$ defined on $G$. The goal is to obtain a closed form expression for $\beta$ and $f_2$ such that $\hat{\eta} = X\hat{\beta} + \hat{f}_2$. We first assume that the solution uniquely exists for each $Y_U$, that is, there exists a unique $R$ with $\hat{\eta}(Y_U) = RY_{Y_U}$. Now apply (2.8) to get that $(X'X)^{-1}\hat{\beta}(Y_U) = X'(Y_{Y_U} - \hat{f}_2(Y_U))$ and $\hat{f}_2(Y_U) = S(Y_{Y_U} - X\hat{\beta}(Y_U))$, which yields

$$\hat{\beta}(Y_U) = (X'(I-S)X)^{-1}X'(I-S)Y_{Y_U}$$

and

$$\hat{f}_2(Y_U) = S(Y_{Y_U} - X\hat{\beta}(Y_U)).$$

Therefore, the Gauss–Seidel algorithm obtains the function estimate $\hat{\eta}(Y_U) = (I-S)X\hat{\beta}(Y_U) + SY_{Y_U} \equiv RY_{Y_U}$. For the fixed point phase, we assume that $\rho(R_{UU}) < 1$ to get that $\hat{Y}_U = \hat{\eta}_U(\hat{Y}_U) = X_U\hat{\beta}(\hat{Y}_U) + \hat{f}_{2_U}(\hat{Y}_U)$. Profiling out $\hat{f}_{2U}$ from $\hat{Y}_U$ and after some algebra, we get that

$$(X'(I-S)X)\hat{\beta} = X'(I-S)Y_{Y_U} = X'(I-S)[\widetilde{Y} + X\hat{\beta}] - X'_L(I - M_{LL})X_L\hat{\beta},$$

where $M_{LL} = S_{LL} + S_{LU}(I - S_{UU})^{-1}S_{UL}$ and $\widetilde{Y}_L = [Y'_L, \widetilde{Y}'_U]'$ with $\widetilde{Y}_U = (I - S_{UU})^{-1}S_{UL}Y_L$. Solving for the coefficient yields the following estimate for both $\beta$ and $f_2$:

(2.11) $\quad \hat{\beta} = (X'_L(I - M_{LL})X_L)^{-1}X'_L(I - M_{LL})Y_L,$

(2.12) $\quad \hat{f}_{2L} = M_{LL}(Y_L - X_L\hat{\beta}) \quad \text{and} \quad \hat{f}_{2U} = M_{UL}(Y_L - X_L\hat{\beta}),$

with $M_{UL} = (I - S_{UU})^{-1}S_{UL}$. This results in $\hat{\eta}_L = (I - M_{LL})X_L\hat{\beta} + M_{LL}Y_L$, which is the natural generalization of the classical semi-parametric modeling result [Hastie and Tibshirani (1990)].[1]

In general, one can apply the Gauss–Seidel algorithm directly to a sequence of smoothers with eigenvalues in $(-1, 1]$. The algorithm is guaranteed to converge to a solution whenever the smoothers are diagonally dominant, or symmetric and positive semidefinite. In other words, one can forgo the first optimization problem and apply the Gauss–Seidel algorithm directly to a sequence of smoothers defined on $X$ and $G$, analogous to the supervised setting [Hastie and Tibshirani (1990)]. For the second step, the solution to the optimization problem technically exists whenever the final matrix $R$ is a transductive smoother. Thus, with well defined smoothers one can always follow this approach to obtain (2.10).

---

[1]In the case of the linear model without the graph term [i.e., $f(X) = X\beta$ or $f(X) = \psi(X)\beta$], we have that (2.11) with $M_{LL} = 0$ results in $\hat{\beta} = \hat{\beta}^{(ls)} = (X'_LX_L)^{-1}X'_LY_L$ (i.e., the supervised ordinary least squares estimate), which is consistent with Culp and Michailidis (2008a).

2.1.1. *Obtaining transductive smoothers for graph views.* We elaborate next on how to obtain the smoother matrix $S$ from an observed graph view $G$. The graph is represented by its weighted $n \times n$ adjacency matrix $A$ defined above, with $A(i,j) \geq 0$. The normalized right stochastic smoother matrix $S$ is then defined by $S = D^{-1}A$, with $D$ being a diagonal matrix containing the row sums (node degrees) of $A$. Notice that the matrices $A$, $D$ and $S$ all emit the necessary partition structure given in (2.9). For example, the weighted similarity adjacency matrix $A$ has four blocks: the $A_{LL}$ block provides weighted links between labeled observations, the $A_{UL}$ and $A_{LU}$ blocks are weighted links between labeled and unlabeled observations, and the $A_{UU}$ block is comprised of weighted links between unlabeled and unlabeled observations. For the document classification example, the weights are defined by document similarity between observations in $L \cup U$.

In the Corollary to Proposition 2 in Culp and Michailidis (2008a), we established that $\rho(S_{UU}) < 1$ whenever $A_{UL} \times \mathbf{1}_L > \vec{0}$ and $A_{UU}$ is irreducible (i.e., these assumptions are sufficient for $S \in \mathcal{A}[G]$). The condition on $A_{UL}$ is interpreted that each unlabeled node has at least one connection to a labeled node. In data involving graphs it should be noted that this condition is difficult to satisfy, especially when the size of the labeled set is small relative to that of the unlabeled set, as illustrated in Figure 1. To account for this, one could start with the observed graph, compute the shortest path distance between nodes and obtain a new complete graph on each component (all nodes are connected to all other nodes within each component), and then, subsequently, "thin" the obtained graph. Notice that these additional steps can circumvent the problem by generalizing to every disconnected component must have a label on it as discussed in Culp and Michailidis (2008a).

The above smoother is generally too simplistic to perform well on real data since there is no tuning parameter $\lambda$. One way to address this is to define the combinatorial Laplacian as $P = D - A$ and then, subsequently, generalize the smoother to $S = (A + \lambda P)^{-1}A$ or the symmetric smoother $S = (I + \lambda P)^{-1}$ in (2.7). Either of these generalizations tend to improve performance over that of the stochastic smoother since the parameter $\lambda$ can be estimated based on the response $Y_L$.

2.1.2. *Obtaining transductive smoothers for feature views.* In the case of feature data, it is fairly straight forward to obtain the transductive smoother from the penalty matrix as given in (2.7). However, additional considerations are made for the case of constructing transductive smoother matrices based on kernel functions, which is the approach primarily used in this work. Specifically, one can construct a similarity matrix $W$, with

$$(2.13) \qquad W_{ij} = K_\gamma(d(x_i, x_j)) \qquad \text{with } i, j \in L \cup U,$$

where $d(\cdot, \cdot)$ is a distance function applied to the vectors containing the data for observations $i$ and $j$, and $K_\gamma$ is a kernel function. The corresponding smoother is then given by $S = (W + \lambda P)^{-1} W$, with $P$ the combinatorial Laplacian of $W$. In the case of $\lambda = 1$, by construction, we have that $S \in \mathcal{A}[X]$ whenever $W_{UL} \times \mathbf{1}_L > \vec{0}$ and $W_{UU}$ is irreducible as with the graph case. However, unlike in the graph case, this condition is typically satisfied in practice when using noncompact kernel functions. On the other hand, performance can improve by introducing a parameter $K$ and constructing adjacency matrix $A$ as the $K$ nearest neighbors defined by similarity associations in $W$ (often referred to as a $K$-NN graph). As a result, the adjacency may require similar modifications as indicated for the observed graph to form the smoother $S = (A + \lambda P)^{-1} A$ (or $S = (I + \lambda P)^{-1}$), where $P$ is now redefined as the combinatorial Laplacian operator defined on the $K$-NN graph $A$.

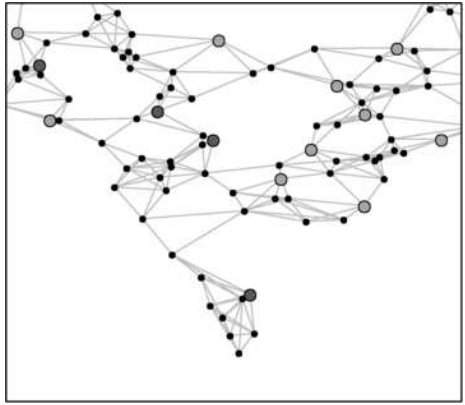

FIG. 1. *A cross-section of an observation graph with both labeled and unlabeled vertices (observations). The labeled vertices consist of either a large dark or light circle, and the unlabeled vertices consist of small black circles. The interest in this example is to illustrate the affect of unlabeled data on the topology of the graph.*

2.1.3. *Parameter estimation.* As noted above, in several instances the proposed framework requires the estimation of several tuning parameters including kernel parameters, nearest neighbor parameters and Lagrangian parameters. In general, the tuning parameters can be broken into two groups: within view and between view ones. The within view parameters denoted by $\tau$ are necessary to construct the penalty matrix $P_{\tau_\ell}$ from either $X$ data sources or graph sources [i.e., $\tau_\ell = (\gamma_\ell, k_\ell)$ to form the $k$-NN graph for view $\ell$]. The between view parameters correspond to the Lagrangian on the penalty matrices for the additive model and are denoted as $\lambda = (\lambda_1, \ldots, \lambda_q)$.

Upon completing the within view parameter estimation step, the goal is to obtain a penalty matrix, $P_\ell$, from each view $\ell$. This problem is treated on a view-by-view basis. For example, in view 1 suppose that it has been established that a random forest learner works particularly well and, similarly, for view 2 a neural net learner is the best performing one. To incorporate this information, we consider a search over the parameter space which finds a transductive smoother $S_\ell$ that predicts similarly to the learner. More precisely, let $\phi_{L\ell}$ and $\phi_{U\ell}$ be the predictions of the procedure applied to view $\ell$ (e.g., a random forest or a neural net) and define the penalty matrix $P_{\tau_\ell} = D_\ell - W_\ell$ using the following criterion:

$$(2.14) \qquad \min_{\tau_\ell} \|\phi_{U_\ell} - (I - S_{UU})^{-1} S_{UL} Y_L\|_2^2 \qquad \text{with } S = D_\ell^{-1} W_\ell.$$

In other words, the goal is to find a value $\gamma_\ell$ and $k_\ell$ such that the solution involving smoother $S$ in (2.10) coincides closely to predictions from learner $\phi$. Notice that the individual smoothers mimic specifically chosen learners within each view, which can result in strong performance in several data applications [refer to the pharmacology data example in Section 3.3, where we observe that the estimate with (2.14) performs quite strongly compared to the state-of-the-art co-training algorithms with random forest].

Upon obtaining the appropriate penalty matrices, one then must deal with the between view parameter estimation problem for $\lambda$ in (2.5). The Lagrangian allows one to simultaneously account for the smoother's contribution for each view. To perform this, we first make use of the fact that the labeled estimate is linear in $Y_L$; that is, $\hat{Y}_L = M_{LL}(\lambda) Y_L$, where $M_{LL}(\lambda) = R_{LL}(\lambda) + R_{LU}(\lambda)(I - R_{UU}(\lambda))^{-1} R_{UL}(\lambda)$ (the notation is modified to indicate the dependance of the smoother matrices on the parameters). In this case, we make use of the standard GCV criterion adjusted to transductive smoothers given by

$$(2.15) \quad \text{tGCV}(M_{LL}(\lambda)) = \min_\lambda \frac{(Y_L - M_{LL}(\lambda) Y_L)'(Y_L - M_{LL}(\lambda) Y_L)}{(1 - \text{tr}(M_{LL}(\lambda))/m)^2}.$$

In practice, the tGCV criteria is optimized simultaneously for each $\lambda_j$ so that adjustments can be made between the views.

REMARK 2. The optimization criterion in (2.15) can also be used to estimate within view parameters when a learner is not available. For example, the parameter $k$ on an observed graph view could be estimated with (2.15). However, in our experience, if a learner is available, then performance can improve by usage of (2.14).

2.1.4. *View interaction terms.* Next we elaborate on the view interaction terms provided in (2.1). We restrict attention to two-way interaction terms that prove most useful in practice. There are three possible types of interactions in the present setting: an interaction between two views comprised of feature data, $f_{12}(X_1, X_2)$, an interaction between two views comprised of observed graphs, $f_{12}(G_1, G_2)$, and, finally, an interaction between a feature and a graph view, $f_{12}(X_1, G_2)$.

In this work we are primarily interested in the case when the views are modeled as transductive kernel or graph smoothers. The interaction term can be defined as a composite graph operation, which can be achieved in various ways. One possibility is to use the intersection, while another the union of the underlying two graphs. The intersection between two graphs $G_i \cap G_j$ with corresponding weighted adjacency matrices $W_i, W_j$ is defined by a new graph whose adjacency matrix is given by $[W_{ij}](u,v) = \sqrt{W_i(u,v)W_j(u,v)}$, while that for their union, $W_{ij} = \frac{W_i + W_j}{2}$.[2] These terms are then processed as additional smoothers in (2.8).

REMARK 3. Another approach for defining interactions is given in Hastie and Tibshirani (1990) where the authors employ restrictions on the $f_{12}(\cdot, \cdot)$ term during estimation. Extensions of this approach to (2.1) could also be considered especially for interactions with nonkernel based terms (e.g., linear terms).

2.2. *Fitting the model in classification.* In classification we assume that the response takes on binary values, $Y_L \in \{0, 1\}$. The goal is to fit a general semi-supervised multi-view model of the form

$$\eta = \alpha + f_1(X) + f_2(G),$$

with $\eta = g(\mu)$, where $g$ is the logit link function. Next we utilize the generalized fixed-point self-training strategy to achieve this objective.

As previously discussed, one must first obtain the training response as $Y_{Y_U} = [Y'_L, Y'_U]'$, where $g(Y_U) \in \mathbb{R}^{|U|}$. For the first step in (2.2), we optimize for $\hat{\eta}$ in

$$(2.16) \qquad \min_{\eta} L(Y_{Y_U}, g^{-1}(\eta)) + \sum_{j=1}^{2} \lambda_j f'_j P_j f_j,$$

---

[2]There are other possibilities for defining the union term, however, we compute it additively [Gould (1998)]. It should be noted that when interaction terms are included in the model, care should be taken to avoid identifiability issues arising from the following situation: $f_{12}(G_1 \cup G_2) \approx f_1(G_1) + f_2(G_2) - f_{12}(G_1 \cap G_2)$.

where the logistic loss function ($L(y,p) = -y\log(p) - (1-y)\log(1-p)$ with $p = g^{-1}(f)$) is used and $P_j$ are the appropriate penalty matrices for $X, G$, respectively. The solution to this problem $\hat{\eta} \equiv \hat{\eta}(Y_U)$ must satisfy

$$\lambda_i P_i \hat{f}_i - (Y_{Y_U} - g^{-1}(\hat{\eta})) = \vec{0}.$$

Employing a Taylor expansion on $g^{-1}$ about $\hat{\eta}^{(k-1)}$ and setting it to zero, we get the following semi-supervised extension of the z-scoring algorithm:

$$\hat{f}_i^{(k)} - S_i^{(k-1)}(\hat{\eta}^{(k-1)} - \hat{f}_i^{(k-1)}) = S_i^{(k-1)} z^{(k-1)},$$

where the smoother is given by $S_i^{(k-1)} = C(W^{(k-1)} + \lambda_i P_i)^{-1} W^{(k-1)}$, the score by $z^{(k-1)} = \hat{\eta}^{k-1} - W^{(k-1)-1}(Y_{Y_U} - g^{-1}(\hat{\eta}^{(k-1)}))$, and the variance function by $W^{(k-1)} = \nabla g^{-1}(\eta)|_{\eta = \eta^{(k-1)}}$. It is easy to see that the above formulation reduces to an application of the Gauss–Seidel algorithm for each $z^{(k)}$. Unlike the regression setting, the smoother for the solution depends on $Y_U$, since $W$ depends on $Y_U$, and hence, we require that there must exist a $R(Y_U)$ such that $\hat{\eta}(Y_U) = R(Y_U) z(Y_U)$. Now, if $W$ is diagonal, then $z_U(Y_U) = \hat{\eta}_U - W_{UU}^{-1}(Y_U - g^{-1}(\hat{\eta}_U))$. From the fixed point step (2.3), we get that $g(\hat{Y}_U) = \hat{\eta}_U(\hat{Y}_U) = z_U$. The iterative self-training algorithm discussed above must be applied in order to obtain this fixed point. The following proposition provides a sufficient condition for the algorithm to converge independent of its initialization (in this case the fixed point must be unique):

PROPOSITION 1. *Assume that the solution $\hat{\eta} = \sum_j \hat{f}_j$ exists and satisfies*

$$\lambda_j P_j \hat{f}_j - (\hat{Y}_{Y_U} - g^{-1}(\hat{\eta})) = 0$$

*for any $Y_U$ such that $g(Y_U) \in \mathbb{R}^{|U|}$. Assume, additionally, that there exists a $\hat{Y}_U$ that satisfies the fixed point solution to (2.3), i.e. $g(\hat{Y}_U) = \hat{\eta}_U(\hat{Y}_U)$. If*

$$\rho\left(\sum_j \sum_{i \neq j} [(I - S_j S_i)^{-1} S_j (I - S_i)]_{UU}\right) < 1,$$

*with $S_i = (\lambda_i P_i + W)^{-1} W$ and $W = \nabla g^{-1}(\eta)|_{\eta = \hat{\eta}}$, then the iterative self-training method converges independent of initialization.*

The condition on the above matrix is not of much practical use, but, nevertheless, it provides a general setting for which the solution to (2.3) uniquely exists.

2.3. *Model selection issues.* Given the multi-view model (2.1) developed above, the next important issue to address is that of model selection, especially in the presence of multiple views and their interactions. We present next a criterion for achieving this objective. To start, for ease of presentation consider a model involving a feature data view $X$ and a graph view $G$. In this work the interest is in view-nested model comparisons; for example, to assess the significance of view interaction term compare

$$(2.17) \qquad \eta = \alpha + f_1(X) + f_2(G) + f_{12}(X, G),$$

$$(2.18) \qquad \eta = \alpha + f_1(X) + f_2(G).$$

An example of such a model comparison is illustrated with the CORA text data (Section 3.2).

The generalized fixed point self training framework provides a natural environment to assess model selection in the multi-view setting. For example, let $\hat{\eta}_1 = R_1 z_1$ and $\hat{\eta}_2 = R_2 z_2$ represent two model estimates. One approach to compare two smoothers is Akaike's Information Criterion (AIC). AIC compares two models by penalizing the loss of the model under consideration with the degrees of freedom of the smoother $(\mathrm{tr}(M_{LL_j}))$, where $M_{LL_j}$ is linear in $z_{L_j}$ for model $j$. Models with lower AIC generally fit better and are less complex than models with larger AICs [Hastie and Tibshirani (1990), Hastie, Tibshirani and Friedman (2001)]. The AIC model comparison for transductive smoothers is formally given by

$$(2.19) \qquad \mathrm{tAIC}(M_{LL_j}) = \frac{2}{m} \mathrm{Loss}(M_{LL_j}) + \frac{2(\mathrm{tr}(M_{LL_j}))}{m},$$

where the $\mathrm{Loss}(M_{LL_j})$ is the $L(Y_L, g^{-1}(M_{LL_j} z_L))$ for some loss function $L$ (for this work we use the logistic loss function). The best model is the one corresponding to the smoother that minimizes tAIC. Also, when optimizing tAIC we preserve the hierarchical constraint, that is, the presence of higher level interaction terms require lower level terms in the model.

2.4. *Implementation issues.* The proposed fitting procedures employ either the Gauss–Seidel algorithm directly, or indirectly through the z-scoring method. However, for regression it is computationally advantageous to employ an iterative backfitting procedure, as opposed to solving the Gauss–Seidel equations directly [Hastie and Tibshirani (1990)]. The main idea behind backfitting is to iteratively smooth each function to the partial residual without that function and subsequently mean center the function. A generalization to the local scoring algorithm is applicable to the generalized additive model setting. Convergence of both algorithms are discussed globally for several possible smoother choices [Buja, Hastie and Tibshirani (1989), Hastie and Tibshirani (1990)]. The transductive local scoring algorithm is presented in Algorithm 1 and can be used to fit all the interesting models under consideration; note that iterative backfitting is a special case of it.

**Algorithm 1** Transductive Local Scoring

1: Initialize vector $\hat{Y}_U^{(0)}$ of size $n$, and select a smoother type for each view (e.g. kernel, linear, spline, etc.) . Input tolerance $\delta > 0$.
2: **for** $k = 1, \ldots$ **do**
3:     **Apply** the local scoring algorithm introduced in Buja, Hastie and Tibshirani (1989) with smoother type specified and response $Y_{\hat{Y}_U^{(k-1)}} = [Y_L', \hat{Y}_U^{(k-1)'}]'$ to determine the estimate $\hat{\eta}^{(k+1)} = [\hat{\eta}_L^{(k+1)'}, \hat{\eta}_U^{(k+1)'}]'$.
4:     **Update** $\hat{Y}^{(k+1)} = g^{-1}(\hat{\eta}^{(k+1)})$ and hence $\hat{Y}_U^{(k+1)} = g^{-1}(\hat{\eta}_U^{(k+1)})$.
5:     Stop if $\| \hat{\eta}_U^{(k+1)} - \hat{\eta}_U^{(k)} \| < \delta$.
6: **end for**

The above algorithm tends to globally converge in practice, but the rate of convergence depends significantly on the choice of $\hat{Y}_U^{(0)}$. The following argument provides a fairly convenient initialization for this procedure. Consider the regression problem in (2.5) with response $Y_{Y_U}$. Previously, we solved for $\hat{\eta}(Y_U) = RY_{Y_U}$ and obtained the fixed point directly. Now, instead, we apply (2.7) to get that $\hat{f}_\ell(Y_U) = S_\ell \hat{\varepsilon}_\ell(Y_U)$, where $\hat{\varepsilon}_\ell(Y_U) = Y_{Y_U} - \hat{\eta}^{(-\ell)}(Y_U)$ is the partial residual, and solve to get that $\hat{\eta}_U(Y_U) = S_{UL_\ell} \hat{\varepsilon}_{L_\ell} + S_{UU_\ell} Y_U + (I - S_{UU_\ell}) \hat{\eta}_U^{(-\ell)}(Y_U)$. Invoke step (2.3) so that $\hat{Y}_U = \hat{\eta}_U(\hat{Y}_U)$ and from this we get that $\hat{Y}_U = (I - S_{UU_\ell})^{-1} S_{UL_\ell} \hat{\varepsilon}_{L_\ell} + \hat{\eta}_U^{(-\ell)}$. Placing this in (2.7) for $\hat{Y}_U$ cancels out and yields $\hat{f}_\ell = M_{\cdot L} \hat{\varepsilon}_{L_\ell}$, where $M_{LL_\ell}$ and $M_{UL_\ell}$ are the smoothers identified in (2.10) for view $\ell$. From this, one can then apply backfitting directly on centered smoothers $M_{LL_\ell}$ with response $Y_L$ to obtain an estimate $\hat{Y}_L^{(0)}$. To obtain the estimate $\hat{Y}_U^{(0)}$, one could predict with the backfitting algorithm using smoothers $M_{UL_\ell}$, response $Y_L$ and previous estimates $\hat{Y}_L^{(0)}$. The initialization $\hat{Y}_U^{(0)}$ tends to be rather close to the solution from the self-training algorithm with centered smoothers $S_\ell$ and response $Y_{\hat{Y}_U}$, and hence, the algorithm converges fairly fast. A similar initialization can be derived for the more general local scoring version.

REMARK 4. When either back-fitting/local scoring are employed for function estimation one must approximate the degrees of freedom used for measurements such as tGCV and tAIC; for example, a common approximate for the denominator of (2.15) is $(1 - [1 + \sum_\ell (\text{tr}(M_\ell) - 1)]/|L|)^2$ [Hastie and Tibshirani (1990)].

**3. Data examples.** To assess the performance and usefulness of the proposed model, we have selected three diverse data examples. In the first ex-ample a synthetic data set comprised of two graph views is examined. In the second example an observed graph is combined with a feature set obtained from text data. In the third example we apply our approach to a pharmaceutical problem.

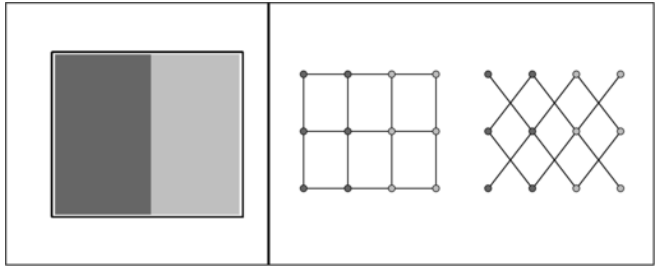

FIG. 2. *The grid of data given on the left has a two-class response represented by the light gray and dark gray points. We consider connecting points using square neighborhoods and diagonal neighborhoods, which correspond to the lattice (right panel).*

3.1. *Graph selection with disjoint lattices.* We consider data from a two-dimensional lattice comprised of 625 nodes on a $25 \times 25$ grid (refer to Figure 2). The example consists of two separate simulated complex response patterns on the graphs as shown in Figure 3 (row 1). For each response configuration, we allow for two different ways to connect neighboring nodes: square and diagonal (refer to Figure 2). Let $L_s$ be the adjacency matrix for the square neighborhood lattice and $L_d$ for the diagonal neighborhood lattice. The following model was considered for each response configuration (checkerboard, or mixed):

$$\eta = \alpha + f_s(L_s) + f_d(L_d). \tag{3.1}$$

The objective was to determine which graphs were important for predicting the response in each configuration using classification accuracy and the tAIC measure.

In the analysis we first sampled 10% of the observations in the lattice to be treated as labeled nodes (cases), while the remaining 90% were treated as unlabeled cases. The weight matrix for each lattice configuration (square, and mixed) was constructed by $W_{ij} = K_\gamma(d_{sp}(i,j))$, with $K_\gamma(d) = e^{-d/\gamma}$, where $d_{sp}(i,j)$ denotes the shortest path from node $i$ to node $j$ on the lattice (note $i, j \in L \cup U$). The penalty matrix for the lattice was given by $P = D - W$, with $D$ the diagonal row sum matrix of $W$. In the estimation of $\eta = \alpha + f_\ell(L_\ell)$ with $\ell \in \{s, d\}$ the smoothers were given by $S_\ell = (W_\ell + P_\ell)^{-1} W_\ell$ and the parameter $\gamma_\ell$ was estimated using the tGCV criterion. For the additive model, $\eta = \alpha + \sum f_\ell$, the penalty matrix $P_\ell$, the kernel matrix $W_\ell$ and the parameter vector $\lambda$ were supplied to the local scoring algorithm as input. However, before proceeding with local scoring, the $\lambda$ parameters were estimated simultaneously using the tGCV criterion analogous to the approach discussed in Hastie and Tibshirani (1990).[3] This process was repeated 50 times. Then, the entire analysis was executed with $Q\%$ of the data treated as labeled ($Q = 20, 30, \ldots, 90$). The average accuracy (second column) and tAIC (third column) for each labeled partition size is reported in Figure 3.

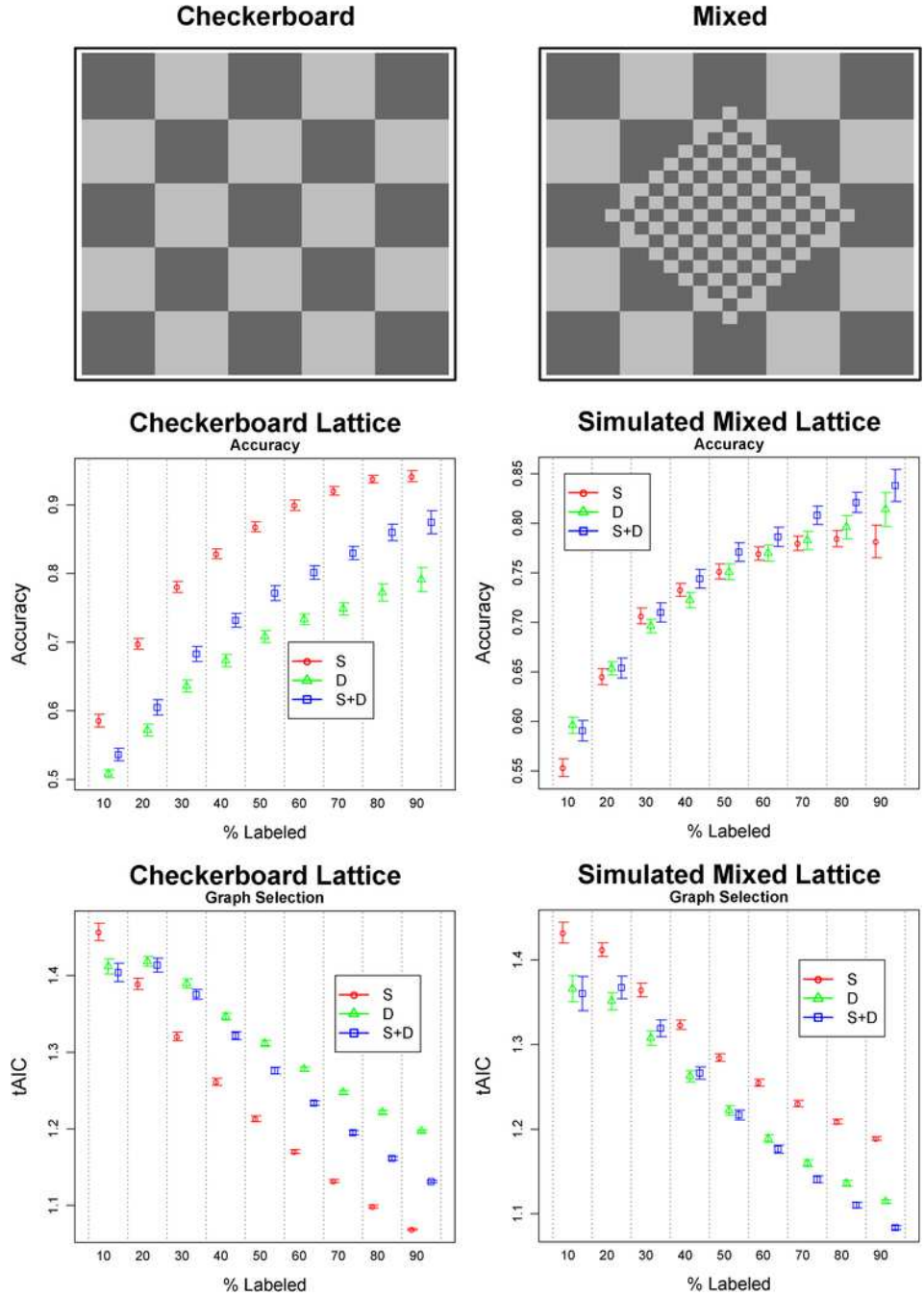

Fig. 3. *The response configurations are presented in row 1. The accuracy for each response configuration is given in row 2, while the tAIC for each response configuration is given in row 3. S denotes square model, D denotes diagonal model, and S+D denotes additive model.*

The accuracy plot for the checkerboard example illustrates that the model with square neighborhoods exhibited the best performance. In the graph selection plot the square neighborhood graph provided the optimal model in terms of smallest tAIC. Therefore, the model suggests that the square neighborhood view fits the data well, in accordance with the underlying checkerboard configuration. For the mixed configuration, the additive model performed marginally better than the other configurations. However, as the size of the labeled data grew, the square and diagonal graphs minimized tAIC.

3.2. *Text analysis.* In our next example we consider 776 documents obtained from the artificial intelligence/machine learning segment of the CORA text data [McCallum et al. (2000)]. The artificial intelligence segment consists of text documents that address the general topics of machine learning, planning, theorem proving, robotics, expert systems and others. The binary response is the indicator that the text document is specifically about machine learning. The first view corresponds to the co-citation network, where the vertices are the labeled and unlabeled text documents, and the edges are the number of times that each pairwise observation agree in citation (co-citation). Specifically, the adjacency matrix is constructed as

$$A_{ij} = \begin{cases} \text{Total \# of documents co-cited with } i, & i = j, \\ \sum I\{i \text{ and } j \text{ cite the same document}\}, & i \neq j. \end{cases}$$

The second view contains 141 carefully parsed words used in the title of each of the text documents (e.g., *learn*, *net*, *theory*, etc.). The text string is a partial match where the first letters of the word in the title must match the variable; for example, if the variable is *net*, then the observation represents a count of any variation of *net*, *nets*, *network*, etc. The following logistic model was used:

(3.2) $\quad \eta = \alpha + f_1(X_{title}) + f_2(G_{cite}) + f_{12}(X_{title}, G_{cite}).$

---

[3]To speed-up computation, transductive backfitting was employed with response $Y_L$ treated as continuous for only the parameter estimation component. Transductive local scoring was used for fitting the actual model in each step.

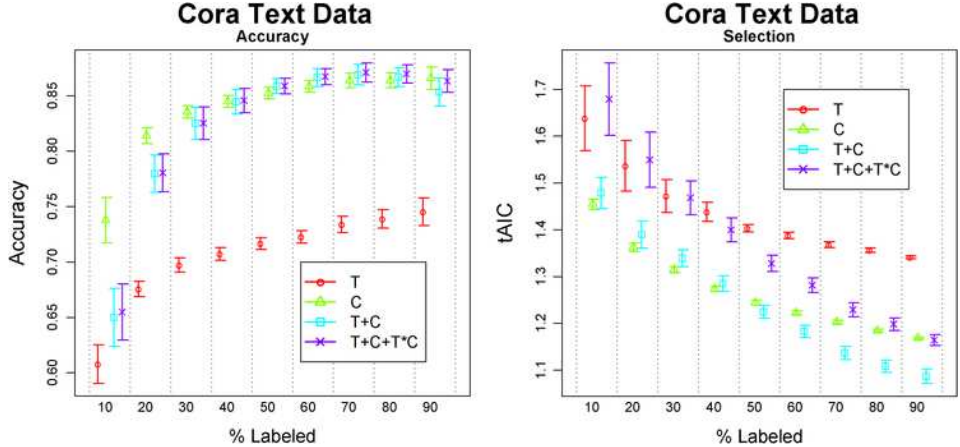

FIG. 4. *(left) The plot provides the average accuracy over 50 samples from applying the multi-viewed model on the cora text data as the amount of labeled data varies from 10% to 90%. (right) The corresponding tAIC plot is provided. T and C denote the models constructed using the word view and the citation view, respectively. T+C is the main effects additive model, and T+C+T\*C is the full model including the interaction between views.*

To compute the interaction term, we employ the intersection operation by defining $W_{12_{ij}} = \sqrt{W_{1_{ij}} * A_{2_{ij}}}$, where $W_1$ is the kernel matrix on the title view and $A_2$ is the co-citation adjacency matrix. The goal is determine the simplest model to adequately predict whether a document is classified as addressing the specific topic of machine learning in the artificial intelligence network.

As before, the percentage of labeled cases was varied from 10% to 90%. The average accuracy results, based on 50 replications, are shown in Figure 4 (left panel) and the tAIC results in Figure 4 (right panel). Cosine dissimilarity was used to construct the distances between observations on the title view and tGCV was used to estimate the parameters. The accuracy results tend to favor the models with both the text and citation information. The tAIC measure indicates that the additive model comprised of the text and citation views without interaction was the minimizer. From this result we select the model $\eta = f_1(X_{title}) + f_2(G_{cite})$ as dominant and drop the interaction term between $X_{title}$ and $G_{cite}$.

Next, we assess the proposed approach against the spectral graph transducer (SGT), first using the $G_{cite}$ view, and then using both the $X_{title}$ and $G_{cite}$ views [Joachims (2003)]. In Figure 5, the SGT(X, G) dominates in the 10–30% configurations and remains competitive for the larger labeled partitions. The SGT(G), however, is only competitive in the 10% configuration.

3.3. *Drug discovery data.* We revisit one of the motivating examples for this work where the observations correspond to compounds which could potentially become drugs. The goal is to assess early in the drug discovery process the potential of a compound to cause an adverse event (AE) or side-effect. Clearly, a compound's AE status is critically important to its success, and as a result, pharmaceutical companies would like to identify these compounds as early as possible in the discovery process. The targeted compounds can then be modified in an attempt to reduce the adverse event status while maintaining their effectiveness, or eliminated from follow-up altogether.

The set of predictors consists of information describing the biological relationship (view 1) between a compound and a particular target and the chemical relationship (view 2) which provides descriptors based on the structure of the compound. In order to obtain the necessary biological information, a therapeutically relevant concentration of the drug is applied to the target, and the inhibition of the target's activity is measured. This view consists of $p_1 = 191$ continuous and noisy variables, each consisting of a carefully chosen target. The chemical predictors are represented by a sparse and binary set of descriptors ($p_2 = 151$), where each descriptor represents a specific substructure in the compound. For the data set under consideration, there are $n = 438$ compounds, of which 92 are known to be associated with a specific AE ($Y = 1$, otherwise $Y = 0$).

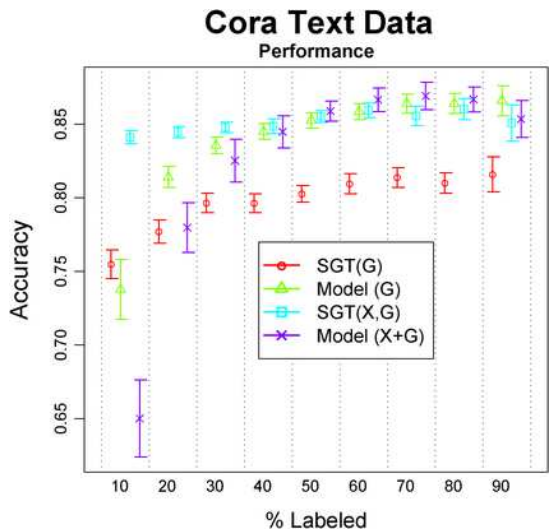

FIG. 5. *The plot provides the accuracy results with 95% confidence bands from applying multi-view model with the spectral graph transducer using the co-citation view only, and the analogous plots with both the title and co-citation views. The amount of labeled data varies from 10% to 90%.*

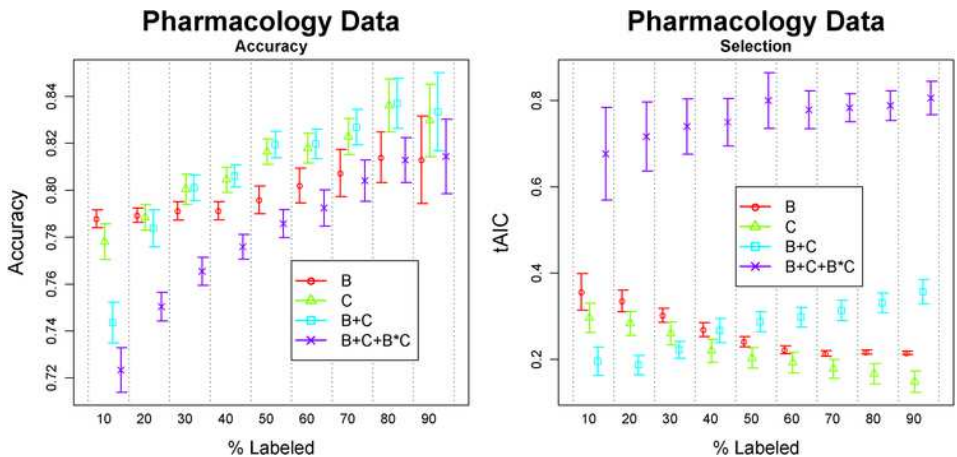

Fig. 6. *(left) The plot provides the average accuracy over 50 samples from applying the multi-view model [recall that random forest were employed in (2.14)] on the pharmacology data as the amount of labeled data varies from* 10% *to* 90%. *(right) The corresponding tAIC plot is provided. B and C denote the models constructed using just the biology view and just the chemistry view, respectively. B+C is the main effects additive model, and B+C+B\*C is the full model including the interaction between views. The 95% confidence bands are provided to assess the precision of the contribution for a particular model.*

In addition to generating a predictive model, scientists are also interested in assessing the importance of each descriptor set. That is, chemical fingerprints (data in view 2) are extremely cheap to obtain, only requiring computation time, whereas the biological information takes more time and money to generate because each compound needs to be assayed. Hence, assessing the importance of both views of information has important resource implications. To determine the appropriate model for this data, the following logistic multi-view model was fitted to the data:

$$\eta = \alpha + f_1(X_{Bio}) + f_2(X_{Chem}) + f_{12}(X_{Bio}, X_{Chem}).$$

The smoothers for each term in the above model were generated by optimizing (2.14), using random forests in each view. The results are shown in Figure 6 for both accuracy and tAIC. In accuracy, the additive model tends to improve over that of the chemistry view only model, but the improvement is not significant. From the tAIC analysis, the additive model with both views seems useful for smaller labeled partitions, but as the %-labeled increases, its utility diminishes to that of the chemistry view only model. This result suggests that the biology view and interactions involving this view are not contributing significantly to the performance of this model.

Next, we wish to assess the proposed modeling approach compared to other multi-view procedures. From the above analysis, the chemistry only view model is all that is necessary, but we use the biology/chemistry model without interaction for comparison with other multi-view techniques. In addition to the multi-view model, a random forest without making a view distinction was fitted to the data [i.e., random forest fit directly to (Bio,Chem)], the co-training procedure discussed in the introduction using random forests as the base learner [Blum and Mitchell (1998)] and the SGT approach was employed on this data. To measure performance, we partition the data into 50 10–90% labeled groups, each with the remainder treated as unlabeled, and applied both accuracy and kappa measures to the unlabeled data. The kappa measure is defined as $(O - E)/(1 - E)$, where $O$ is the observed agreement in the testing confusion matrix and $E$ is the expected agreement [Cohen (1960)]. Values close to 0 represent poor agreement, while values close to 1 represent perfect agreement. Because kappa compares observed agreement to expected agreement, it is helpful for assessing performance for unbalanced data.

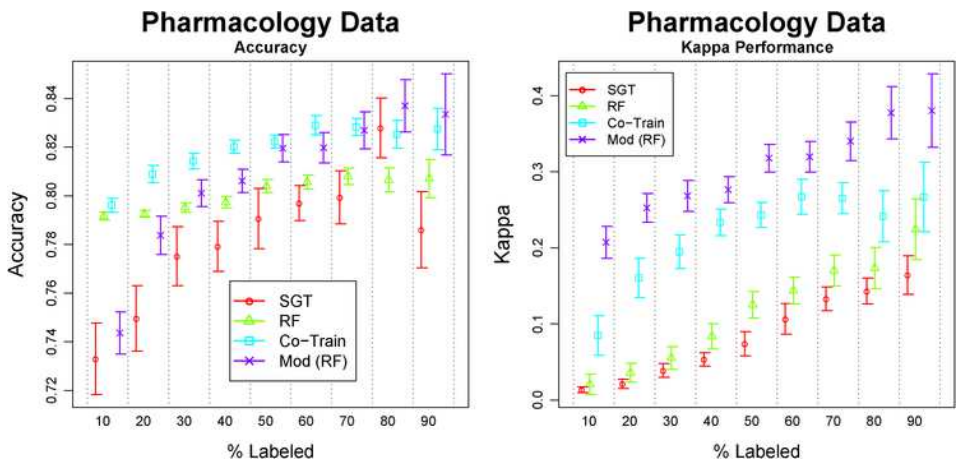

FIG. 7. *(left) The plot provides the accuracy results with 95% confidence bands from applying the multi-view model with random forest, Co-Training with RF, the spectral graph transducer, and the random forest without a view distinction on the pharmacology data as the amount of labeled data varies from* 10% *to* 90%. *(right) The plot provides the analogous results for the Kappa performance measure, which better accounts for unbalanced classes.*

The multi-view model with random forest in (2.14) and the co-training procedure are quite competitive in both the accuracy and kappa measures (refer to Figure 7). The results based on kappa reveal that co-training is somewhat more conservative than the multi-view model (i.e., tends to predict several observations as not having an AE), and therefore, the multi-view model exhibits a strong performance in kappa with a slight deterioration in accuracy. The supervised random forest applied without view distinction and the SGT performed poorly with respect to both measures for this data set.

**4. Concluding remarks.** In this paper we developed a modeling framework suitable for analyzing multi-view data. Its main features are: (i) the generalized fixed point framework to fit semi-supervised additive models to both observed graph and feature views, (ii) mechanisms to perform view selection and incorporate view interactions and (iii) data-driven tuning parameter estimation. The proposed framework and subsequent developments provide a marked departure from the original co-training algorithms into a data analysis setting by allowing view interactions and selections.

A topic of future study is the ability to assess variable contribution within and between views. In this setting, the individual feature views are constructed from variables and, therefore, the contribution of a particular view depends on that of the underlying variables. On the other hand, a variable's contribution may occur at the view level such as in an interaction. Other interesting issues of study involve inference, testing and transductive covariance estimation. The process of predicting new observations (as opposed to retraining, which is the current process) is also under investigation. The difficulty of this problem, often referred to as inductive learning, is noted in several references [Culp and Michailidis (2008), Krishnapuram et al. (2005), Zhu, Ghahramani and Lafferty (2003), Wang and Zhang (2006), Zhu (2007)] and is worthy of its own investigation.

**5. Proof of Proposition 1.** Next we provide proof for Proposition 1.

PROPOSITION 1. *Assume that the solution $\hat{\eta} = \sum_j \hat{f}_j$ exists and satisfies*

$$\lambda_j P_j \hat{f}_j - (\hat{Y}_{Y_U} - g^{-1}(\hat{\eta})) = 0$$

*for any $Y_U$ such that $g(Y_U) \in \mathbb{R}^{|U|}$. Assume, additionally, that there exists a $\hat{Y}_U$ that satisfies the fixed point solution to (2.3), that is, $g(\hat{Y}_U) = \eta_U(\hat{Y}_U)$. If*

$$\rho\bigg(\sum_j \sum_{i \neq j} [(I - S_j S_i)^{-1} S_j (I - S_i)]_{UU}\bigg) < 1,$$

*with $S_i = (\lambda_i P_i + W)^{-1} W$ and $W = \nabla g^{-1}(\eta)|_{\eta = \hat{\eta}}$, then the iterative self-training method converges independent of initialization.*

PROOF. By assumption, we have that

$$\lambda_j P_j (\hat{f}_j^{(k)} - \hat{f}_j) = (Y_{\hat{Y}_U^{(k-1)}} - Y_{\hat{Y}_U}) - (g^{-1}(\hat{\eta}^{(k)}) - g^{-1}(\hat{\eta})),$$

where $\hat{\eta} = \hat{\alpha} + \sum_j \hat{f}_j$ and $\hat{\eta}^{(k)} = \hat{\alpha}^{(k)} + \sum_j \hat{f}_j^{(k)}$. We also have that

$$g^{-1}(\hat{\eta}^{(k)}) = g^{-1}(\hat{\eta}) + W(\hat{\eta}^{(k)} - \hat{\eta}) + O(1)n,$$

with $W = \nabla g^{-1}(\eta)|_{\eta=\hat{\eta}}$. Putting these together, we get that

$$\lambda_j P_j(\hat{f}_j^{(k)} - \hat{f}_j) = (Y_{\hat{Y}_U^{(k-1)}} - Y_{\hat{Y}_U}) - W(\hat{\eta}^{(k)} - \hat{\eta}) + O(1),$$

and hence,

$$\hat{f}_i^{(k)} - \hat{f}_i = S_i\left[W^{-1}(Y_{\hat{Y}_U^{(k-1)}} - Y_{\hat{Y}_U}) - \sum_{i \neq j}(\hat{f}_j^{(k)} - \hat{f}_j)\right] + O(1),$$

with $S_i = (W + \lambda_i P_i)^{-1} W$. After some algebra and solving for a common term,

$$(I - S_i S_j)(\hat{f}_i^{(k)} - \hat{f}_i) = S_i(I - S_j)W^{-1}(Y_{\hat{Y}_U^{(k-1)}} - Y_{\hat{Y}_U}) + O(1), \qquad i \neq j.$$

Assume $W$ is a multiple of $I$ and define $R_i = (I - S_i S_j)^{-1} S_j(I - S_j)$, then

$$\hat{f}_{U_i}^{(k)} - \hat{f}_{U_i} = R_{UU_i}(\hat{\eta}_U^{(k-1)} - \hat{\eta}_U) + O(1).$$

From this, we can cycle the above statements with $i = 1, \ldots, q$ for $\ell = 1, \ldots, k$ to get that

$$\hat{f}_{L_i}^{(k)} - \hat{f}_{L_i} = R_{LU_i}\left(\sum_{j=1}^{q} R_{UU_j}\right)^{k-1}(\hat{\eta}_U^{(0)} - \hat{\eta}_U) + O(1),$$

$$\hat{f}_{U_i}^{(k)} - \hat{f}_{U_i} = R_{UU_i}\left(\sum_{j=1}^{q} R_{UU_j}\right)^{k-1}(\hat{\eta}_U^{(0)} - \hat{\eta}_U) + O(1).$$

Therefore, if $[\sum_{j=1}^{q} R_{UU_j}]^k \to 0$, then convergence of the algorithm is guaranteed. The actual initialization $\hat{\eta}^{(0)}$ is of no consequence, therefore, the convergence is independent of initialization. $\square$

**Acknowledgments.** The authors would like to thank the area Editor Michael Stein, the AE and three anonymous referees for their useful comments and suggestions.

M. Culp  
Department of Statistics  
P.O. Box 6330  
West Virginia University  
Morgantown, West Virginia 26506  
USA  
E-mail: [mculp@stat.wvu.edu](mculp@stat.wvu.edu)

G. Michailidis  
Department of Statistics  
439 West Hall  
1085 South University  
University of Michigan  
Ann Arbor, Michigan 48109-1107  
USA  
E-mail: [gmichail@umich.edu](gmichail@umich.edu)

K. Johnson  
Pfizer Global Research & Development  
Ann Arbor, Michigan 48105  
USA  
E-mail: [Kjell.Johnson@pfizer.com](Kjell.Johnson@pfizer.com)